\documentstyle[12pt]{article}

\begin{document}
\topmargin -35pt
\oddsidemargin 5mm

\newcommand {\beq}{\begin{eqnarray}}
\newcommand {\eeq}{\end{eqnarray}}
\newcommand {\non}{\nonumber\\}
\newcommand {\eq}[1]{\label {eq.#1}}
\newcommand {\defeq}{\stackrel{\rm def}{=}}
\newcommand {\gto}{\stackrel{g}{\to}}
\newcommand {\hto}{\stackrel{h}{\to}}
\newcommand {\1}[1]{\frac{1}{#1}}
\newcommand {\2}[1]{\frac{i}{#1}}
\newcommand {\th}{\theta}
\newcommand {\thb}{\bar{\theta}}
\newcommand {\ps}{\psi}
\newcommand {\psb}{\bar{\psi}}
\newcommand {\ph}{\varphi}
\newcommand {\phs}[1]{\varphi^{*#1}}
\newcommand {\sig}{\sigma}
\newcommand {\sigb}{\bar{\sigma}}
\newcommand {\Ph}{\Phi}
\newcommand {\Phd}{\Phi^{\dagger}}
\newcommand {\Sig}{\Sigma}
\newcommand {\Phm}{{\mit\Phi}}
\newcommand {\eps}{\varepsilon}
\newcommand {\del}{\partial}
\newcommand {\dagg}{^{\dagger}}
\newcommand {\pri}{^{\prime}}
\newcommand {\prip}{^{\prime\prime}}
\newcommand {\pripp}{^{\prime\prime\prime}}
\newcommand {\prippp}{^{\prime\prime\prime\prime}}
\newcommand {\pripppp}{^{\prime\prime\prime\prime\prime}}
\newcommand {\delb}{\bar{\partial}}
\newcommand {\zb}{\bar{z}}
\newcommand {\mub}{\bar{\mu}}
\newcommand {\nub}{\bar{\nu}}
\newcommand {\lam}{\lambda}
\newcommand {\lamb}{\bar{\lambda}}
\newcommand {\kap}{\kappa}
\newcommand {\kapb}{\bar{\kappa}}
\newcommand {\xib}{\bar{\xi}}
\newcommand {\ep}{\epsilon}
\newcommand {\epb}{\bar{\epsilon}}
\newcommand {\Ga}{\Gamma}
\newcommand {\rhob}{\bar{\rho}}
\newcommand {\etab}{\bar{\eta}}
\newcommand {\chib}{\bar{\chi}}
\newcommand {\tht}{\tilde{\th}}
\newcommand {\zbasis}[1]{\del/\del z^{#1}}
\newcommand {\zbbasis}[1]{\del/\del \bar{z}^{#1}}
\newcommand {\vecv}{\vec{v}^{\, \prime}}
\newcommand {\vecvd}{\vec{v}^{\, \prime \dagger}}
\newcommand {\vecvs}{\vec{v}^{\, \prime *}}
\newcommand {\alpht}{\tilde{\alpha}}
\newcommand {\xipd}{\xi^{\prime\dagger}}
\newcommand {\pris}{^{\prime *}}
\newcommand {\prid}{^{\prime \dagger}}
\newcommand {\Jto}{\stackrel{J}{\to}}
\newcommand {\vprid}{v^{\prime 2}}
\newcommand {\vpriq}{v^{\prime 4}}
\newcommand {\vt}{\tilde{v}}
\newcommand {\vecvt}{\vec{\tilde{v}}}
\newcommand {\vecpht}{\vec{\tilde{\phi}}}
\newcommand {\pht}{\tilde{\phi}}
\newcommand {\goto}{\stackrel{g_0}{\to}}
\newcommand {\tr}{{\rm tr}\,}
\newcommand {\GC}{G^{\bf C}}
\newcommand {\HC}{H^{\bf C}}
\newcommand{\vs}[1]{\vspace{#1 mm}}
\newcommand{\hs}[1]{\hspace{#1 mm}}
\newcommand{\al}{\alpha}
\newcommand{\be}{\beta}

\setcounter{page}{0}

\begin{titlepage}

\begin{flushright}
TIT/HEP-463\\
{\tt hep-th/0101166}\\
January 2001
\end{flushright}
\bigskip

\begin{center}
{\LARGE\bf
A Note on Supersymmetric WZW term\\ 
in Four Dimensions
}
\vs{10}

\bigskip
{\renewcommand{\thefootnote}{\fnsymbol{footnote}}
\large\bf Muneto Nitta\footnote{
E-mail: nitta@th.phys.titech.ac.jp}
}

\setcounter{footnote}{0}
\bigskip

{\small \it
Department of Physics, Tokyo Institute of Technology,\\
Oh-okayama, Meguro, Tokyo 152-8551, Japan\\
}

\end{center}
\bigskip

\begin{abstract}
We reconsider the supersymmetric Wess-Zumino-Witten (SWZW) term 
in four dimensions. 
It has been known that 
the manifestly supersymmetric form of the SWZW term includes 
derivative terms on auxiliary fields, the highest components 
of chiral superfields, and then 
we cannot eliminate them by their equations of motion. 
We discuss a possibility for 
the elimination of such derivative terms 
by adding total derivative terms.  
Although the most of derivative terms can be eliminated as in this way, 
we find that all the derivative terms can be canceled,  
{\it if and only if} an anomalous term in SWZW term vanishes.  
As a byproduct, we find the {\it first} example of 
a higher derivative term free from such a problem.  
 
\end{abstract}

\end{titlepage}

When a global symmetry is spontaneously broken down 
to its subgroup in four dimensions,  
there appear massless Nambu-Goldstone (NG) bosons. 
Low-energy effective field theories for these massless bosons 
are nonlinear sigma models, which can be constructed by 
the nonlinear realization method~\cite{CCWZ}. 
These effective Lagrangian include two derivative terms, 
and then are dominant at the low-energy scale. 
Corrections to effective Lagrangian are higher derivative terms. 
Especially anomalies for the global symmetry can be 
reproduced at the low-energy scale 
by the Wess-Zumino-Witten (WZW) term, which 
is a four derivative term~\cite{WZ,Wi}. 
 
If spontaneously symmetry breaking occurs 
in $N=1$ supersymmetric field theories in four dimensions, 
massless NG bosons, with additional bosons and 
fermionic superpartner,  
constitute massless chiral superfields~\cite{SNG}. 
Low-energy effective field theories are described by 
supersymmetric nonlinear sigma models~\cite{Zu},  
which can be constructed by a supersymmetric extension of 
the nonlinear realization method~\cite{BKMU}.
In these models, anomalies of the global symmetry 
can be reproduced at the low-energy scale by 
the supersymmetric extension of WZW (SWZW) term~\cite{CL,NR,CG}. 
(For anomalies in supersymmetric gauge theories, 
see Ref.~\cite{Su} and references therein.)  
Manifestly supersymmetric form of the SWZW term 
has been firstly constructed by 
Nemeschansky and Rohm~\cite{NR}, 
in which bosonic parts consist of 
the bosonic WZW term and an additional four derivative term. 
Four derivative terms of the Skyrme type also has been 
considered in Ref.~\cite{BNS}. 
(See Ref.~\cite{higher}, for higher derivative terms 
in supersymmetric standard models.)
Low-energy effective Lagrangian of supersymmetric QCD 
including the SWZW term has been discussed in Ref.~\cite{Ma},  
but only bosonic part has been discussed there. 

It was, however, pointed out in Ref.~\cite{BNS} 
that higher derivative terms including SWZW term 
have a serious problem:  
There are terms with spacetime derivatives 
acting on auxiliary field, namely the highest components 
of chiral superfields, and then  
we cannot eliminate them by their equations of motion. 
Auxiliary fields must propagate!  
To overcome this problem, Gates and his collaborators 
has suggested a new form of the SWZW term 
by doubling the chiral superfields 
to chiral and complex linear superfields~\cite{Ga}. 
However this doubling has not been justified yet 
in the framework of $N=1$ supersymmetric 
low-energy effective field theories,   
since there is no Nambu-Goldstone theorem for 
additional superfields.\footnote{
In recent works of Gates and his collaborators, 
the SWZW term are further studied without 
the doubling~\cite{Ga2}. 
Although the component level analysis is not completed, 
there is possibility that these models are also free from 
the problem of the auxiliary field.
} 

In this note we reconsider the SWZW term 
in supersymmetric nonlinear sigma models  
consisting of chiral superfields only. 
There remain a possibility for 
the elimination of derivative terms on auxiliary fields  
by adding total derivative terms.  
We discuss this possibility and find that 
most derivative terms can be eliminated in this way, 
but there remain some non-vanishing derivative terms. 
The condition for disappearance of these terms 
is equivalent to the condition 
on the vanishing of an anomalous term. 
Then this problem cannot be avoided 
in supersymmetric nonlinear sigma models with the SWZW term, 
which correctly reproduce anomalies.
We find, however, the {\it first} example of 
higher derivative terms free from this problem.\footnote{
In the nonlinear elecro-dynamics, 
there exists an example of higher derivative terms  
free from this problem~\cite{KLRT}.
}    
After we recapitulate 
the bosonic WZW term and supersymmetric nonlinear sigma models 
without the WZW term, 
we discuss the SWZW term. 

\bigskip
Before discussing supersymmetric field theories, 
we recall fundamental facts of the WZW term in 
bosonic field theories to fix notations.  
The Lagrangian of nonlinear sigma models is 
\beq
 {\cal L}= \1{2} g_{ij}(\phi) 
 \del_{\mu}\phi^i \del^{\mu}\phi^j. 
\eeq
where $\phi^i(x)$ are NG fields for 
a global symmetry breaking of $G$ to $H$. 
The number of NG fields is $i=1,\cdots,\dim(G/H)$, 
and $g_{ij}(\phi)$ is a metric of coset space $M = G/H$. 
The WZW term in bosonic nonlinear sigma models can be 
written as the integration of a closed $5$-form $\omega$ 
on $M$, $d \omega =0$,  
over the five dimensional extension of the spacetime~\cite{Wi}:  
\beq
 S_{\rm WZW} = c \int_5 \omega (\hat\phi^i(x,t)), 
 \hs{10} \hat\phi^i(x,0) = \phi^i(x), 
\eeq
where $\hat\phi^i(x,t)$ is an extension of fields 
to the five dimensions, 
and $t$ is an extra coordinate. From the Poincare's lemma, 
$\omega$ can be written locally by 
a potential $4$-form $\lam$ as
$\omega = d \lam$. 
Then, due to the Stokes' theorem, the WZW term becomes      
\beq
 S_{\rm WZW} = c \int_4 \lam(\phi(x)) 
 = c \int d^4 x \eps^{\mu\nu\rho\sig}\lam_{ijkl}(\phi)
   \del_{\mu}\phi^i \del_{\nu}\phi^j
   \del_{\rho}\phi^k \del_{\sig}\phi^l,  \label{WZ}
\eeq
which is the integrated WZW term. 

In the case of the supersymmetric nonlinear sigma models, 
NG fields sit in chiral superfields 
$\Phi(x,\th,\thb)$, 
satisfying $\bar D_{\dot\alpha} \Phi(x,\th,\thb) =0$. 
A chiral superfield can be expanded by 
the Grassmannian coordinate $\theta$ and $\thb$ as~\cite{WB} 
\beq
 \Phi(y,\th) &=& A^i(y) 
  + \sqrt 2 \th \psi^i(y) + \th\th F^i(y) \non
 \Phi(x,\th,\thb) &=& A^i(x)+i\th\sig^{\mu}\thb \del_{\mu}A^i(x) 
   + \1{4} \th\th\thb\thb \Box A^i(x) + \sqrt 2 \th \psi^i(x) \non
  && -\2{\sqrt 2} \th\th\del_{\mu} \psi^i(x) \sig^{\mu} \thb
   + \th\th F^i(x),
\eeq
where $y^{\mu} = x^{\mu} + i \th \sig^{\mu}\thb$. 
The Lagrangian of supersymmetric nonlinear sigma models 
can be written as~\cite{Zu} 
\beq
 \int d^4 \th K(\Phi,\Phi\dagg) 
 = g_{ij^*}(A,A^*) \del_{\mu}A^i \del^{\mu}A^{*j} + \cdots, 
 \label{snlsm}
\eeq
where dots denote fermionic terms. 
Using the K\"{a}hler potential $K(\Phi,\Phi\dagg)$,
the metric is defined by 
\beq
 g_{ij^*}(A,A^*) = {\del^2 K(A,A^*) \over \del A^i \del A^{*j}}.
\eeq
Then target manifolds of supersymmetric nonlinear sigma model 
are K\"{a}hler manifolds~\cite{Zu}.
Note that, in the calculation of Eq.~(\ref{snlsm}), 
we must eliminate auxiliary field $F^i$ 
of chiral superfields by their equations of motion
\beq
 F^i = {\Gamma^i}_{jk}(A,A^*) \psi^j\psi^k , \label{el.F}
\eeq
where ${\Gamma^i}_{jk}$ is the connection on 
the K\"{a}hler manifold.

Let us discuss the supersymmetric analog of the WZW term. 
The fundamental object for WZW term still be 
a closed $5$-form $\omega(A,A^*)$ on the K\"{a}hler manifold.
Situations are, however, slightly different from bosonic theories. 
The closed condition on $\omega$ in the K\"{a}hler manifold is     
\beq
 d \omega = (\del + \bar\del) \omega = 0, 
\eeq
where $\del$ and $\bar\del$ are 
exterior derivatives with respect to 
holomorphic and anti-holomorphic 
coordinates $A^i$ and $A^{*i}$, respectively. 
(For a review of geometries of K\"{a}hler manifolds, see, e.g., 
a textbook~\cite{Na}.)
In this case, $\omega$ can be locally written as    
\beq
 \omega = \del \bar\del \beta, 
\eeq 
where $\beta(A,A^*)$ is a potential $(2,1)$-form:
$\beta = \beta_{ijk^*} dA^i \wedge dA^j \wedge dA^{*k}$.~\footnote{
Therefore 
$\beta$ is a $(2,1)$-form and $\omega$ is a $(3,2)$-form.
}
The manifestly supersymmetric expression of SWZW term is 
given in Ref.~\cite{NR} by 
\beq
 S_{\rm SWZW} 
 = ic \int d^4 x d^4 \th \left[
  \beta_{ijk^*}(\Phi,\Phi\dagg) D^{\alpha}\Phi^i 
  {\sig^{\mu}}_{\alpha\dot\beta}\del_{\mu}\Phi^j
  \bar D^{\dot\beta}\Phi^{\dagger k} + {\rm c.c.}\right]. 
   \label{SWZW}
\eeq
It was shown in Ref.~\cite{NR} that if we put $F^i=0$ by hand 
in the bosonic part ${\cal L}_{\rm boson}$, it becomes
\beq
 {\cal L}_{\rm boson}|_{F^i=0} 
 &=& - 4 i \eps^{\mu\nu\rho\sig}\lam_{ijk^*l^*}(A,A^*)
   \del_{\mu}A^i \del_{\nu}A^j 
   \del_{\rho}A^{*k} \del_{\sig}A^{*l}\non
 && - 8 \chi_{ijk^*l^*}(A,A^*)
   \del_{\mu}A^i \del^{\mu}A^{*k} 
   \del_{\nu}A^j \del^{\nu}A^{*l}. 
  \label{bosonic} 
\eeq
Here $(2,2)$-forms $\lam$ and $\chi$ are defined by 
\beq
 && \lam_{ijk^*l^*} 
 = \beta_{ijk^*,l^*} - \bar \beta_{k^*l^*i,j}, \hs{10}
 \lam = \bar \del \beta - \del \bar \beta , \\
 && \chi_{ijk^*l^*} 
 = \beta_{ijk^*,l^*} + \bar \beta_{k^*l^*i,j},\hs{10}
 \chi = \bar \del \beta + \del \bar \beta, 
\eeq
where ``,'' denote differentiations with respect to scalar fields. 
The first term proportional to $\lam$ in Eq.~(\ref{bosonic}) 
is just the bosonic part of the (integrated) SWZW term, 
which is the complex extension of Eq.~(\ref{WZ}); 
on the other hand, the second term proportional to $\chi$
is a non-anomalous $G$-invariant term. 
The $5$-form $\omega$ can be written by these forms as 
\beq
 \omega = \del \lam = \del \chi.
\eeq

Although we have put $F^i=0$ by hand 
in the calculation of Eq.~(\ref{bosonic}), 
we have to eliminate auxiliary field $F^i$ 
by their equations of motion for a complete calculation, 
as done in Eq.~(\ref{el.F}).  
There are, however, terms with spacetime derivatives acting on $F^i$, 
and we cannot eliminate them by their equations of motion.  
There remains a possibility to cancel 
the derivative terms by adding total derivative terms, 
which do not change the action, 
or by partial integrals.  
In this note we discuss this possibility.  
Any term of type $X(A,\psi) \del_{\mu} F$, 
in which $X(A,\psi)$ does not include $F^i$,
can be replaced with 
$-F \del_{\mu} X(A,\psi)$ by adding a total derivative term 
$-\del_{\mu} \left[X(A,\psi) F\right]$. 
Dangerous terms are $F \del_{\mu} F$ terms.  
(There are no $\del_{\mu} F \del_{\nu} F$ terms as shown below.) 

Let us calculate $F \del_{\mu} F$ terms for concreteness.
The pull-back factors $\del_{\mu} \Phi^i$ and $D_{\al} \Phi^i$ 
in Eq.~(\ref{SWZW}) can be written by component fields as 
\beq
 \del_{\mu} \Phi^i 
 &=& \del_{\mu}A^i + i\th\sig^{\nu}\thb \del_{\mu}\del_{\nu}A^i 
   + \1{4} \th\th\thb\thb \Box \del_{\mu} A^i 
   + \sqrt 2 \th \del_{\mu}\psi^i \non
 && -\2{\sqrt 2} \th\th \del_{\mu}\del_{\nu} \psi^i \sig^{\nu} \thb
   + \th\th \underline{\del_{\mu}F^i}, \label{del-Phi}\\
 D_{\al} \Phi^i 
 &=& \sqrt 2 \psi^i_{\al} + 2 \th_{\al}F^i 
 + 2i (\sig^{\mu}\thb)_{\al}\del_{\mu} A^i 
 + \th_{\al}\thb\thb \Box A^i 
 - \2{\sqrt 2} (\th\sig_{\mu}\thb) 
   (\sig^{\mu}\sigb^{\nu}\del_{\nu}\psi^i)_{\al} \non
 && + i\sqrt 2 (\sig_{\mu}\thb)_{\al} \th \del_{\mu}\psi^i 
 - \1{2\sqrt 2} \th\th\thb\thb \Box \psi^i_{\al} 
 + i \th\th (\sig^{\mu}\thb)_{\al} \underline{\del_{\mu}F^i}, 
  \label{D-Phi}
\eeq
where the argument $x$ of component fields in the right-hand-side 
are implicit. 
The last terms associated with underlines 
in both equations are just dangerous $\del_{\mu} F$ terms.

On the other hand, 
the tensor part $\beta_{ijk^*}(\Phi,\Phi\dagg)$ 
includes no $\del_{\mu} F$ terms. 
To expand it by $\th$ and $\thb$, 
we note the Taylor expansion of $\beta$ around $\Phi=0$,   
\beq
 \beta_{ijk^*}(\Phi,\Phi\dagg)
 = \sum_{N,M=0}^{\infty} \1{N!M!} 
 \beta_{ijk^*,i_1 \cdots i_N j_1^* \cdots j_M^*}|_0 
 \Phi^{i_1} \cdots \Phi^{i_N} 
 \Phi^{\dagger j_1}\cdots  \Phi^{\dagger j_M}, \label{Taylor}
\eeq 
where coefficients are evaluated at $\Phi=0$. 
If we define  
\beq
 K_{N,M}(\Phi,\Phi\dagg) = \Phi^{i_1} \cdots \Phi^{i_N}
 \Phi^{\dagger j_1}\cdots  \Phi^{\dagger j_M}, \nonumber
\eeq  
its expansion by $\th$ and $\thb$ can be calculated, to yield 
\beq
 && \hs{5} K_{N,M}(\Phi,\Phi\dagg) \non
 &&= K_{N,M}(A,A^*) + \th(\cdots) + \thb(\cdots)
  + \th\th \left[F^i\del_i K_{N,M}(A,A^*) + \cdots \right] \non
 && + \thb\thb \left[F^{*i}\del_{i^*} K_{N,M}(A,A^*) + \cdots \right]
  + \th \sig^{\mu}\thb (\cdots) \non
 && + \th\th\thb \left[ \sqrt 2 \bar\psi^i F^j 
                \del_{i^*}\del_j K_{N,M}(A,A^*) + \cdots\right]
  + \thb\thb\th \left[ \sqrt 2 \psi^i F^{*j} 
                \del_i\del_{j^*} K_{N,M}(A,A^*) + \cdots\right] \non
 && + \th\th\thb\thb \left[ 
       F^{*i}F^j \del_{i^*}\del_j K_{N,M}(A,A^*) 
  - \1{2} F^{*i} \psi^j\psi^k \del_{i^*}\del_j\del_k 
                                      K_{N,M}(A,A^*) \right.\non
  &&\left. \hs{10}
  - \1{2} F^i \psb^j\psb^k \del_i\del_{j^*}\del_{k^*} K_{N,M}(A,A^*) 
  + \cdots\right],
\eeq
where ``$\cdots$'' denote terms not including $F^i$. 
(For our purpose, it is sufficient to calculate terms 
concerning with $F^i$.) 
By summing up the Taylor expansion (\ref{Taylor}) again, 
we obtain the expansion of $\beta$ by $\th$ and $\thb$, given by 
\beq
 && \hs{5} \beta_{ijk^*}(\Phi,\Phi\dagg) \non
 &&= \beta_{ijk^*} + \th(\cdots) + \thb(\cdots)
  + \th\th \left[F^l \beta_{ijk^*,l} + \cdots \right]  
  + \thb\thb \left[F^{*l} \beta_{ijk^*,l^*} + \cdots \right]
  + \th \sig^{\mu}\thb (\cdots) \non
 && + \th\th\thb \left[ \sqrt 2 \bar\psi^m F^l 
             \beta_{ijk^*,lm^*} + \cdots\right]
  + \thb\thb\th \left[ \sqrt 2 \psi^l F^{*m} 
             \beta_{ijk^*,lm^*} + \cdots\right] \non
 && + \th\th\thb\thb \left[ 
       F^{*m}F^l \beta_{ijk^*,lm^*} 
     - \1{2} F^{*n} \psi^l\psi^m \beta_{ijk^*,lmn^*}
     - \1{2} F^l \psb^m\psb^n \beta_{ijk^*,lm^*n^*}
     + \cdots\right], \label{beta-exp}
\eeq
where all $\beta$ terms at the right-hand-side are evaluated at 
bosonic fields $(A,A^*)$, 
and ``$\cdots$'' again denote terms not including $F^i$.

By calculating the $\th\th\thb\thb$ term in 
Eq.~(\ref{SWZW}) comprising of 
(\ref{beta-exp}), (\ref{D-Phi}), (\ref{del-Phi}) 
and the complex conjugation of (\ref{D-Phi}),  
the $F \del_{\mu} F$ terms can be obtained, to give 
\beq
 && 2 \th\th\thb\thb \left[   
     \{ i (F^i\del_{\mu} F^{*k} + F^{*k}\del_{\mu} F^i) 
       \del^{\mu}A^j\beta_{ijk^*}(A,A^*) + \mbox{c.c.} \} \right.\non
 &&\left. \hs{10}
    - \{F^{*k} \del_{\mu} F^j (\psb^l \sigb^{\mu} \psi^i) 
       \beta_{ij[k^*,l^*]}(A,A^*) + \mbox{c.c.} \} \right], 
  \label{F-del-F}
\eeq
where $[\;,\;]$ represent anti-symmetrization of indices. 
($T_{[ij]} = T_{ij} - T_{ji}$.) 
Since $\del_{\mu} F$ terms in Eqs.~(\ref{D-Phi}) and (\ref{del-Phi}) 
are proportional to $\th\th$, 
there are no $\del_{\mu} F\del_{\nu} F$ terms in 
the $\th\th\thb\thb$ term in the product. 
The first two terms in Eq.~(\ref{F-del-F}) can be written as  
$\del_{\mu} (F^{*k} F^i) \del^{\mu}A^j \beta_{ijk^*}+ \mbox{c.c.}$,  
and then they can be replaced with  
$- F^{*k} F^i \del_{\mu} (\del^{\mu}A^j \beta_{ijk^*})
+ \mbox{c.c.}$ 
by adding total derivative terms. 
On the other hand, by adding a total derivative term, 
the last two terms can be replaced with 
\beq
 && - 2 \th\th\thb\thb  
    F^{*k} \del_{\mu} F^j (\psb^l \sigb^{\mu} \psi^i) 
    \left[ \beta_{ij[k^*,l^*]}(A,A^*) 
    - \bar \beta_{k^*l^*[i,j]}(A,A^*)\right], 
\eeq
and this cannot be eliminated 
by adding any total derivative term. 
Hence the necessary and sufficient condition 
for the last term to vanish is 
\beq
 \beta_{ij[k^*,l^*]} = \bar \beta_{k^*l^*[i,j]}, \hs{10}
 \bar\del \beta = \del \bar\beta, 
\eeq
which gives 
\beq
 \lam = 0 , \hs{10} 
 \chi = 2 \bar\del \beta = 2 \del \bar \beta. \label{cond.}
\eeq
We thus have found that 
the auxiliary fields $F^i$ can be free from 
the derivative problem {\it if and only if} 
the anomalous term $\lam$ disappears. 
In this condition, however, 
the non-anomalous higher derivative term 
of the $(2,2)$-form $\chi$ in Eq.~(\ref{bosonic}) survives, 
which is the {\it first} example of a higher derivative term 
free from propagating auxiliary fields. From Eq.~(\ref{cond.}) 
and $\del^2=\bar\del^2=0$, $\chi$ is a closed $(2,2)$-form: 
\beq
 d \chi = (\del + \bar\del) \chi = 0.
\eeq

We have discussed the manifestly supersymmetric extension 
of the WZW term, and have found that 
the auxiliary field problem cannot be avoided 
unless an anomalous term vanish. 
We have found the first example of 
a higher derivative term in which auxiliary field 
is free from derivative terms.  

This term would play a role in the construction of 
manifestly supersymmetric soliton, like the Skyrmion. 
A manifestly supersymmetric extension of 
the original Skyrme term is also an interesting task. 
A classification of possible higher derivative terms 
would be needed for supersymmetric 
chiral perturbation theories. 
We hope that these studies contribute to 
developments of supersymmetric field theories.  

\section*{Acknowledgements}
The author is grateful to Yutaka Ohkouchi for  
the arguments in the early stage of this work.
He also would like to appreciate S.~J.~Gates,~Jr. for 
pointing out Ref.~\cite{KLRT} and explaining his recent works. 
His work is supported in part 
by JSPS Research Fellowships.


\end{document}